\documentstyle[12pt,epsfig, graphicx,caption,subcaption]{article}
\begin{document}
\def\scri{\unitlength=1.00mm
\thinlines
\begin{picture}(3.5,2.5)(3,3.8)
\put(4.9,5.12){\makebox(0,0)[cc]{$\cal J$}}
\bezier{20}(6.27,5.87)(3.93,4.60)(4.23,5.73)
\end{picture}}

\begin{center}

{\Large WHAT ARE EXTREMAL KERR}

\

\

{\Large KILLING VECTORS UP TO?}

\vspace{18mm}

{\large Jan E. \AA man}\footnote{ja@fysik.su.se}

\vspace{7mm}

{\large Ingemar Bengtsson}\footnote{ibeng@fysik.su.se}

\vspace{7mm} 

{\large Helgi Freyr R\'unarsson}\footnote{helgi.runarsson@gmail.com}

\vspace{8mm}

{\sl Fysikum, Stockholms Universitet,} 

{\sl S-106 91 Stockholm, Sweden}

\vspace{20mm}

{\bf Abstract:}

\end{center}

\

\noindent In the extremal Kerr spacetime the horizon Killing vector field is 
null on a timelike hypersurface crossing the horizon at a fixed latitude, and 
spacelike on both sides of the horizon in the equatorial plane. We explain in 
some detail how this behaviour is consistent with the existence of 
timelike Killing vectors everywhere off the horizon, and how 
it arises in a limit from the Kerr spacetime where there is a 
similar hypersurface strictly outside the horizon.

\newpage

{\bf 1. Introduction}

\vspace{5mm}

\noindent The Kerr solution is one of the most important exact solutions 
ever obtained in physics, and has been thoroughly studied \cite{ON}. 
Its event horizon is ruled by a Killing vector field whose causal character 
changes from timelike to spacelike across the horizon. This is the horizon 
Killing field. The difference from a static black hole is that the horizon 
Killing field becomes null also on a timelike hypersurface surrounding 
the horizon. Still the existence of a Killing vector field which is timelike 
in a region around the event horizon is important for instance in setting up 
curved spacetime quantum field theory there \cite{Frolov}.  

A special case of the Kerr solution is its extremal limit, which has attracted 
attention recently for reasons that have more to do with quantum gravity than with 
astrophysics \cite{CFT}. Its event horizon is again ruled by a horizon Killing 
field, but this Killing field becomes null on a timelike hypersurface crossing 
the horizon at some fixed latitude. In the equatorial plane one finds that the 
horizon Killing vector field is spacelike except at the horizon itself \cite{Racz, 
Amsel}. This behaviour is very different from that encountered in non-extreme 
Kerr, and also in the spherically 
symmetric extremal black holes that we are familiar with. We found it puzzling. 
Given that the exterior of the extremal Kerr black hole is stationary, what is the 
Killing vector field that is timelike all the way down to the horizon? And how 
can we understand the behaviour of the timelike hypersurface where the horizon 
Killing field goes null as a limiting case of the corresponding hypersurface in 
the non-extreme Kerr black hole? 

It turns out that the first question is posed incorrectly. We will rephrase it, and then 
give its answer, in section 2. The second question is not obviously well posed since 
the notion of limits of spacetimes is a subtle one \cite{Geroch}. Nevertheless it has a simple 
and intuitively appealing answer, that we give in section 3. Section 4 contains some 
comments on the near horizon geometry of extremal Kerr. In section 5 we state our conclusions. 

Before we begin, we remind the reader that the Kerr metric in Boyer-Lindquist 
coordinates is 

\begin{equation} ds^2 = - \frac{\Delta - a^2\sin^2{\theta}}{\rho^2}dt^2 
- \frac{4mar\sin^2{\theta}}{\rho^2}dtd\phi + \frac{\rho^2}{\Delta}dr^2 + \rho^2
d\theta^2 + \frac{A\sin^2{\theta}}{\rho^2}d\phi^2 \ , \end{equation} 

\noindent where 

\begin{equation} \rho^2 \equiv r^2 + a^2\cos^2{\theta} \ , \hspace{10mm} 
A \equiv (r^2 + a^2)^2 - a^2\Delta \sin^2{\theta} \ , \end{equation}

\begin{equation} \Delta \equiv r^2 - 2mr + a^2 = (r-r_+)(r-r_-) \ , 
\hspace{6mm} r_\pm = m 
\pm \sqrt{m^2-a^2} \ . \end{equation}

\noindent This coordinate system divides the Kerr spacetime into three regions: 

\begin{equation} \mbox{I}: \ r_+ < r < \infty \ , \hspace{6mm} \mbox{II}: \ 
r_- < r < r_+ \ , \hspace{6mm} \mbox{III}: - \infty < r < r_- \  . \end{equation}

\noindent Region I is the physically relevant exterior, region II lies between 
the two Killing horizons at $\Delta = 0$, and region III contains an unphysical 
asymptotic region. The mass of the black hole is $m$, its angular momentum $J = am$, 
and its event horizon is at $r = r_+$. The extremal case is obtained by setting 
$a = m$, in which case there is no region II since the two horizons coincide. The 
$a < m$ case is believed to 
describe actually existing black holes (in the approximation that these can be 
regarded as isolated), while the case $a > m$ is a nakedly singular spacetime.  

A general Killing vector field of the Kerr solution (whether extremal or not) 
is given by a linear combination 

\begin{equation} \xi = \partial_t + b\partial_\phi \ , \end{equation}

\noindent where $b$ is a constant. Its norm is 

\begin{equation} ||\partial_t + b\partial_\phi ||^2 = \frac{1}{\rho^2}\left[ 
-\Delta + \sin^2{\theta}(a^2-4marb + Ab^2)\right] \ . \end{equation}

\noindent This can vanish for all $\theta$ only if $\Delta = 0$, that is 
on one of the horizons. It vanishes on the outer horizon $r = r_+$ if and only if 

\begin{equation} b = \frac{a}{2mr_+} \ . \end{equation}

\noindent This particular linear combination defines the horizon Killing 
vector field $\xi_{\rm hor}$. One finds indeed 

\begin{equation} || \xi_{\rm hor}||^2 = - \frac{r-r_+}{4m^2\rho^2r_+^2}f \ , 
\label{norm} \end{equation}

\noindent where

\begin{eqnarray} f = f(r,\theta) = - 4m^2a^2\sin^2{\theta}(r-r_+) + \hspace{40mm} 
\nonumber \\ 
\label{f} \\
+ \left( 4m^2r_+^2 - a^2(r^2+2mr+a^2)\sin^2{\theta} + a^4\sin^4{\theta}\right) 
(r-r_-)  \ . 
\nonumber \end{eqnarray}

\noindent We will soon analyze these expressions in detail. 

\vspace{10mm}

{\bf 2. Timelike Killing vector fields in extreme Kerr}

\vspace{5mm}

\noindent In the extreme case $m = a = r_+ = r_-$ the expression for the norm 
of the horizon Killing field is easily factorized:

\begin{equation} || \xi_{\rm hor}||^2 = - \frac{(r-m)^2}{4m^2\rho^2}f_1f_2 \ , 
\end{equation}

\begin{equation} f_1 = \left( m\sin^2{\theta} + (r+m)\sin{\theta} - 2m \right) 
\ , \end{equation}

\begin{equation} f_2 = \left( m\sin^2{\theta} - (r+m)\sin{\theta} - 2m\right) 
\ . \end{equation}

\noindent There is a double root at the horizon ($r = m$). The other roots describe 
two timelike hypersurfaces where the horizon Killing field is null, one of them 
at negative values of $r$. We are interested in the root defined by $f_2 = 0$. 
This describes a hypersurface crossing the horizon at  

\begin{equation} \sin{\theta} = \sqrt{3}-1 \ , \end{equation}

\noindent namely at $\theta \approx 47^\circ$. 

In the equatorial plane $\theta = \pi/2$ 
the horizon Killing field is spacelike on both sides of the horizon. Still there must 
be timelike Killing vectors there. This is made obvious by studying the Killing bivector 

\begin{equation} K^{ab} = \partial_t^a\partial_\phi^b - \partial_t^b\partial_\phi^a \ . 
\end{equation}

\noindent Its norm is 

\begin{equation} \frac{1}{2}K_{ab}K^{ab} = - \Delta \sin^2{\theta} \ . \end{equation}

\noindent We see that the Killing vectors span a timelike 2-plane everywhere, except 
on the horizon ($\Delta = 0$) and on the axes ($\sin{\theta} = 0$). Another way of making 
this obvious is to observe that the Bardeen vector field, which describes the motion of 
a zero angular momentum observer, is timelike outside the horizon. At each point this 
vector field is a linear combination of the Killing vectors. See exercise 33.3 in ref. 
\cite{MTW}. But the Bardeen vector field is not in itself a Killing vector field, since 
the linear combination is spacetime dependent. Also it would seem that a Killing vector 
field that is timelike just 
outside the horizon must be null on the horizon, and the horizon vector field---which 
is spacelike outside the horizon at least on the equatorial plane---is 
the only Killing vector field that does become null on the horizon. So how can there be 
a Killing vector field which is timelike in a region just outside the horizon?

The resolution turns out to be that there is no such Killing vector field. Nevertheless 
there are timelike Killing vectors at all points off the horizon. To simplify the 
calculations let us study the situation in the equatorial plane $\theta = \pi/2$ where 
the norm of a general Killing vector field is

\begin{equation} ||\partial_t + b\partial_\phi ||^2 = \frac{1}{r}\left[ -r+2m 
-4m^2b + b^2(r^3+m^2r + 2m^3)\right] \ . \end{equation}

\noindent We will set $m = 1$ for simplicity. Then one finds that the norm vanishes 
if 

\begin{equation} b = \frac{2\pm r(r-1)}{r^3+r+2} = \left\{ \begin{array}{l} b_+ = 
\frac{1}{r+1} \\ \\ b_- = \frac{2-r}{r^2-r+2} \ , \end{array} \right.  \end{equation}

\noindent It is however more interesting to see where a Killing field with a fixed 
value of $b$ is null. If $b = 0$ this happens at $r = 2$. This is where the ergosphere 
intersects the equatorial plane. If $b \neq 0$ we find that the norm vanishes if 
and only if 

\begin{equation} r^3 + \frac{b^2-1}{b^2}r + \frac{2(b-1)^2}{b^2} = 0 \ . \end{equation}

\noindent The latter polynomial has three roots, 

\begin{equation} r = \frac{1-b}{b} \end{equation}

\begin{equation} r = \frac{1}{2b}\left( b-1 \pm \sqrt{(1-b)(7b+1)}\right) \ . \end{equation}

\begin{figure}
\centerline{ \hbox{
                \epsfig{figure=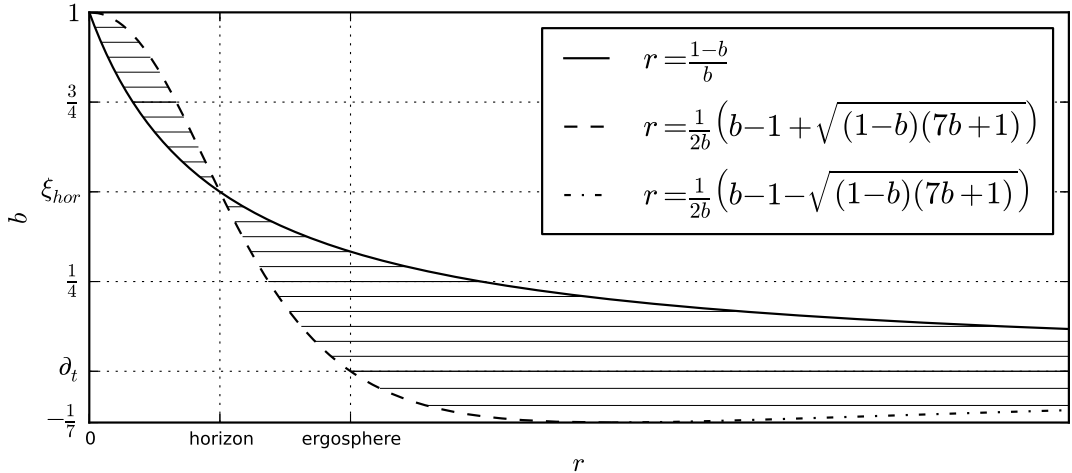,width=145mm}}}
\caption{\small{This picture shows the region of the equatorial plane 
where a given Killing vector field $\partial_t + b\partial_\phi$ is 
timelike, for various values of $b$.}} 
\label{fig:onion}
\end{figure}

\noindent However, we are interested only in real roots, and primarily in positive 
real roots. We therefore confine ourselves to $b$ lying between $-1/7$ and $+1$. 
The result is shown in Fig. \ref{fig:onion}. 

For $b = 1/2$ (the horizon Killing field) the only real root is at $r = 1$, that is at 
the horizon itself. For smaller values of $b$ there is an annulus in the equatorial 
plane where the Killing vector field is timelike. For larger values of $b$ this 
annulus lies inside the horizon. There is one special Killing vector field, namely 
$\partial_t$, which becomes timelike at $r = 2$ (the ergosphere) and ``escapes'' to 
infinity while staying timelike all the way out. For our purposes the point is that 
although every point in the equatorial plane belongs to one of the annuli, none 
of them actually extend all the way to the horizon. This is the answer to our 
first question.

\vspace{10mm}

{\bf 3. The velocity-of-light surface}

\vspace{5mm}

\noindent The timelike hypersurface where the horizon vector field goes null is called 
the velocity-of-light surface in ref. \cite{Amsel}. Possibly this is an unfortunate 
terminology (since other hypersurfaces go under the same name), but we will adopt it 
here. To understand its limiting behaviour as $a \rightarrow m$ we have to 
analyze the full Kerr case. Going back to eqs. (\ref{norm}-\ref{f}), we observe 
that the horizon Killing field is null both at the horizon ($r = r_+$) and at the 
hypersurfaces defined by $f = f(r, \theta) = 0$. Such a hypersurface can be described 
either as a function $r = r (\theta )$ or as a function $\theta = \theta (r)$. 
Indeed one finds 

\begin{equation} \sin{\theta} = \sqrt{\frac{h - \sqrt{h^2-g}}{2a^2\Delta}} \ , 
\end{equation}

\noindent where 

\begin{equation} h = (r^2+a^2)^2 - 4m^2\left( a^2+2r_+(r-m)\right) \ , \end{equation}

\begin{equation} g = 16m^2\Delta^2(2mr_+-a^2) \ . \end{equation}  

\noindent To solve for $r$ we must find the roots of a third order polynomial. This 
can be done using Mathematica, but the explicit expressions do not look pleasant. 
Still we did use numerical solutions for $r$ in constructing our plots. 

\begin{figure}
    \vspace{-10mm}
    \begin{center}
        \begin{subfigure}[H]{0.5\textwidth}
                \centering
                \includegraphics[height=.225\textheight]{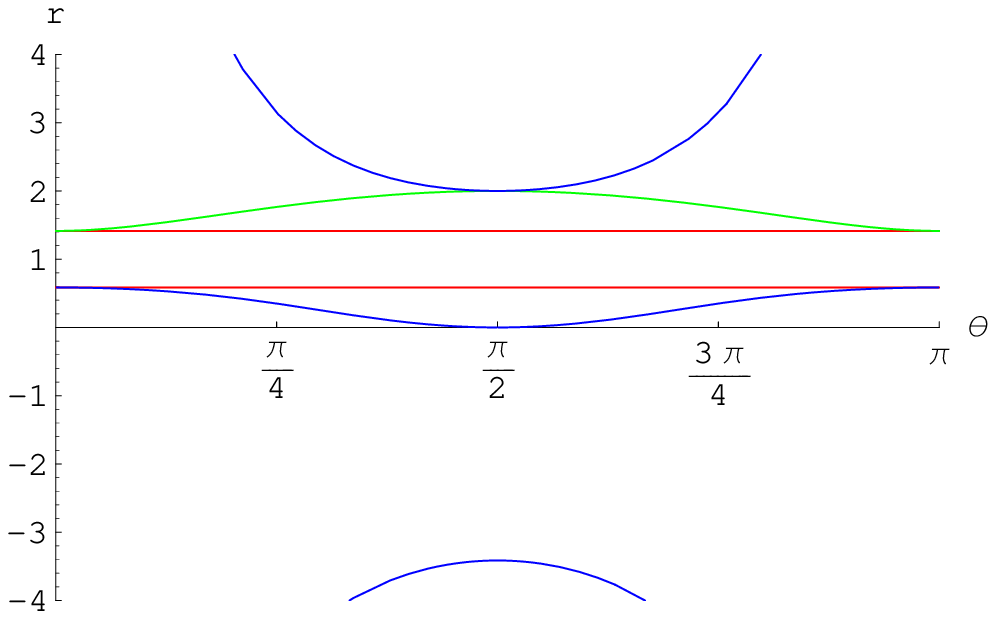}
        \end{subfigure}%
        ~
        \begin{subfigure}[H]{0.5\textwidth}
                \centering
                \includegraphics[height=.325\textheight]{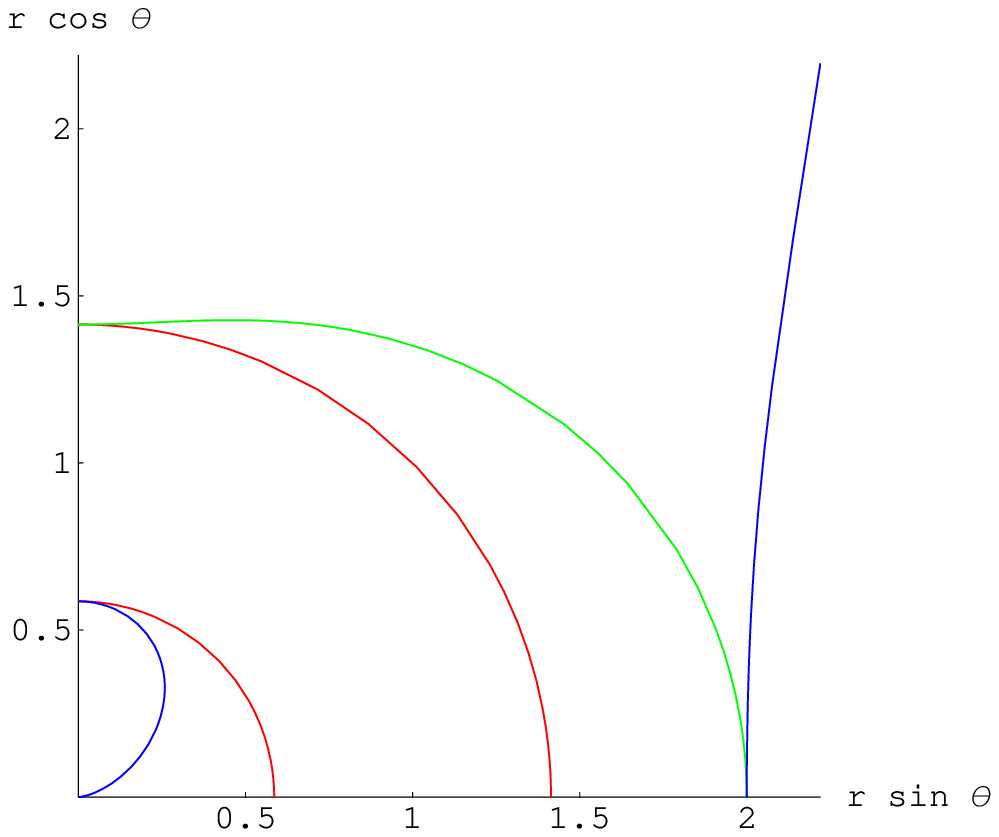}
        \end{subfigure}
        
         \begin{subfigure}[H]{0.5\textwidth}
                \centering
                \includegraphics[height=.225\textheight]{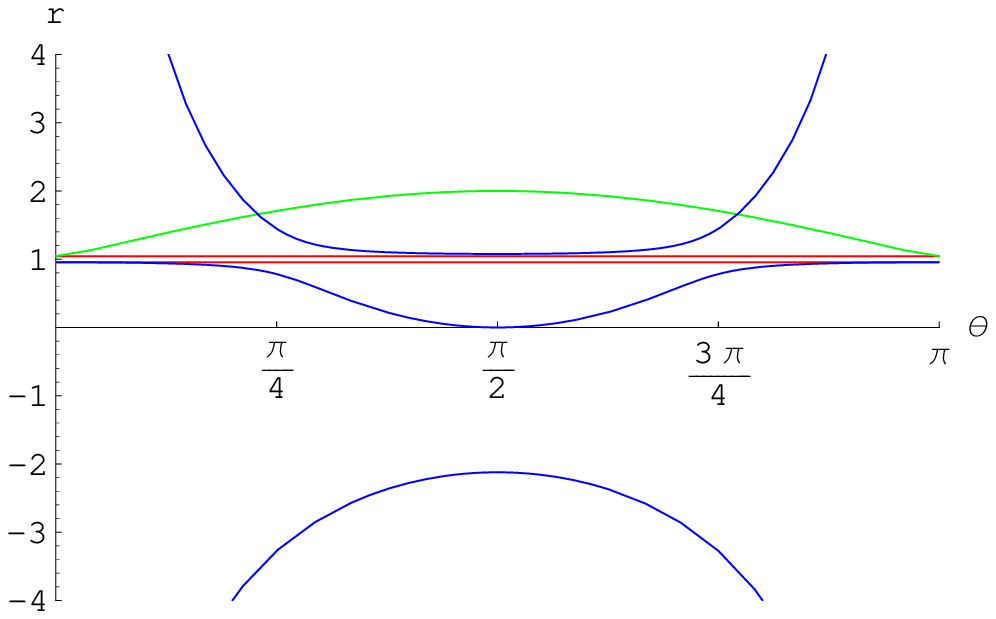}
        \end{subfigure}%
        ~
        \begin{subfigure}[H]{0.5\textwidth}
                \centering
                \includegraphics[height=.325\textheight]{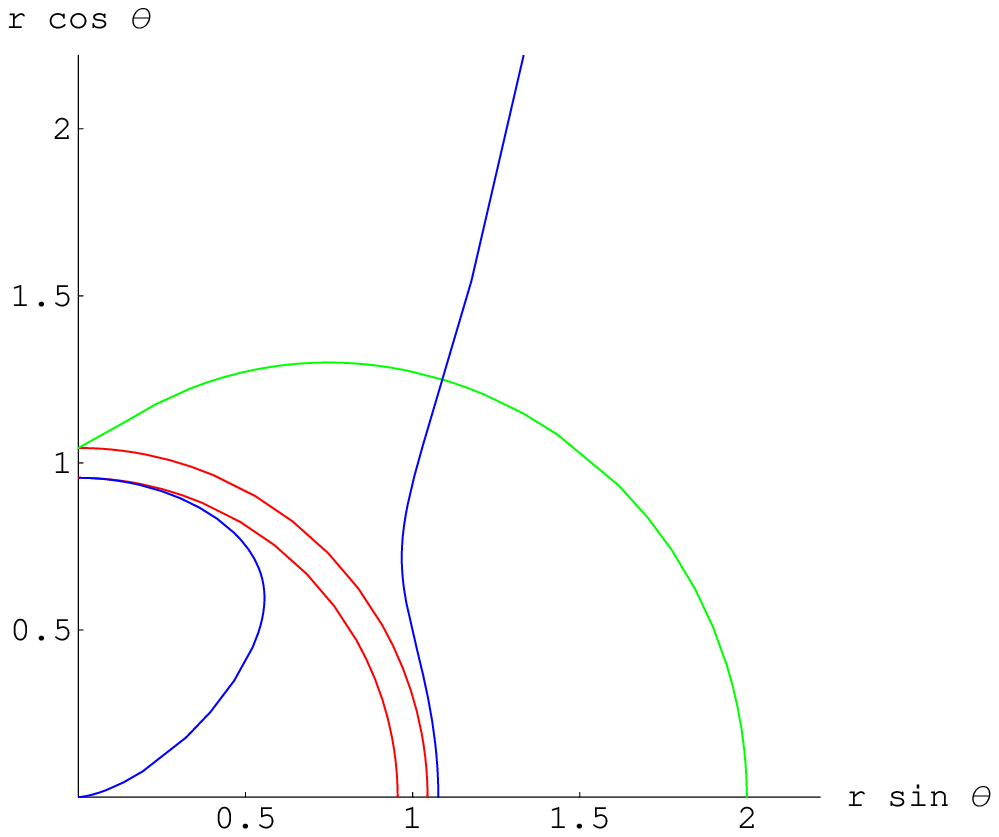}
        \end{subfigure}

         \begin{subfigure}[H]{0.5\textwidth}
                \centering
                \includegraphics[height=.225\textheight]{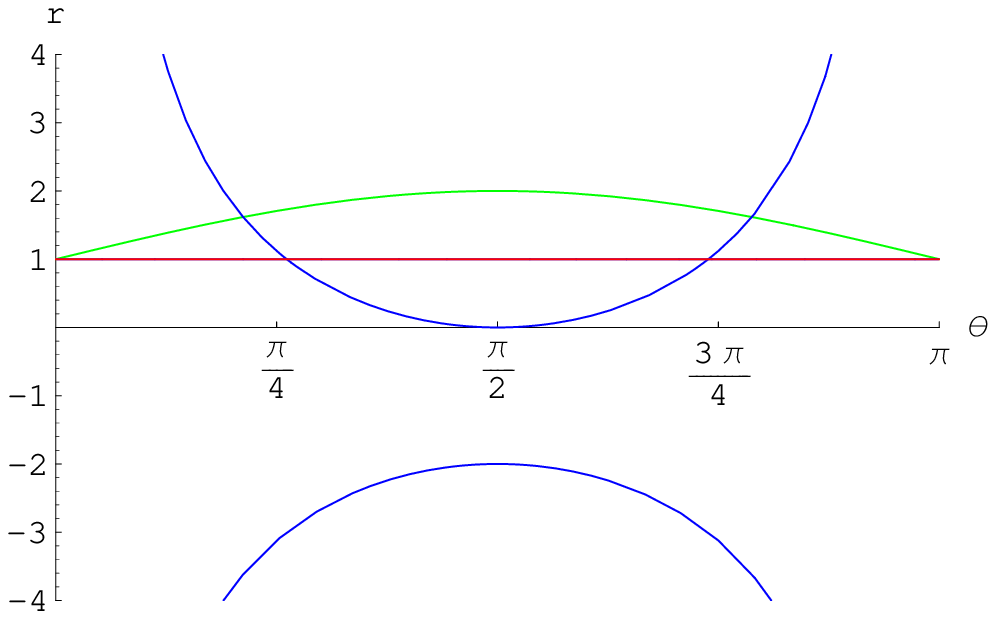}
        \end{subfigure}%
        ~
        \begin{subfigure}[H]{0.5\textwidth}
                \centering
                \includegraphics[height=.325\textheight]{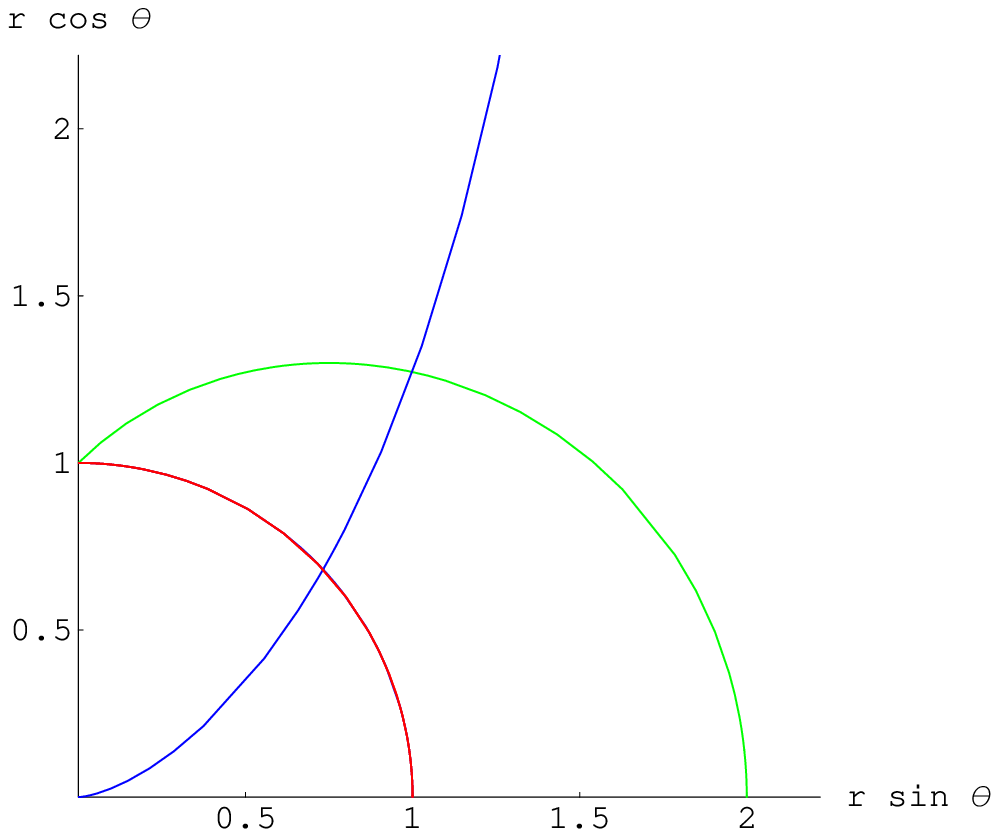}
        \end{subfigure}
        \caption{The velocity-of-light surfaces for different values of $\frac{a}{m}$. See the text for explanations.}\label{fig:2}
    \end{center}
\end{figure}

In Fig. \ref{fig:2} we show the results in the $r$-$\theta$-plane. The left hand 
column treats $r$ and $\theta$ as spanning a plane, and we see three velocity-of-light 
surfaces. The right hand column treats $r$ and $\theta$ as polar coordinates. This 
has some intuitive appeal, but it misses the admittedly unphysical region with negative 
values of $r$. Three different values of $a/m$ are illustrated. From the 
top down they are  
$a/m = \sqrt{2(\sqrt{2}-1)} \approx 0.91$, which is the value of the parameter 
for which the velocity-of-light surface touches the ergosphere, $a/m = 0.999$, and 
$a/m = 1$, which is the extremal case. The outer pair of velocity-of-light 
surfaces always touch a horizon at the poles. One velocity-of-light surface  
always touches the singularity, which sits at $(r,\theta) = (0,\pi/2)$. In 
the plots we have set $m = 1$, the outer and inner 
horizons are red, the ergosphere is 
green, and the velocity-of-light surfaces are blue. 
 
\begin{figure}[h]
\centerline{ \hbox{
                \epsfig{figure=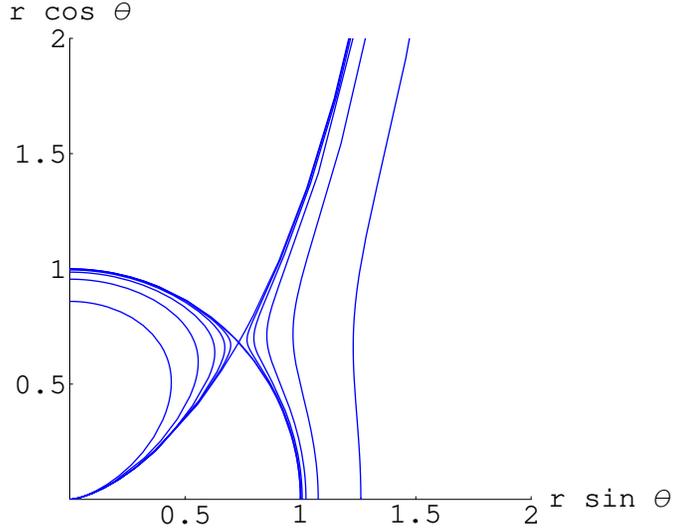,width=90mm}}}
\caption{\small{The velocity-of-light surfaces for $m=1$ and six different values 
of $a/m$: 0.99, 0.999, 0.9999, 0.99999, 0.999999, and 1, together with 
the event horizon of the extremal black hole.}} 
\label{fig:curves}
\end{figure}

In Fig. \ref{fig:curves} we see clearly how two 
velocity-of-light surfaces in the Kerr solution merge and form a single hypersurface 
crossing the horizon in the limit as $a/m$ tends to 1. This is the answer to our 
second question.

\vspace{10mm}

{\bf 4. The near horizon geometry}

\vspace{5mm}

\noindent For the extremal black hole there is only one velocity-of-light surface, 
and it has 
such a simple description that we are able to report its intrinsic curvature. 
In particular, let $R$ be its scalar curvature and $S_{ab} = R_{ab} - 1/3g_{ab}$ 
the traceless part of its Ricci tensor. Using CLASSI \cite{CLASSI} to perform 
the calculation we obtain 

\begin{equation} R = - \frac{s^3(s+1)(12s^8 - 86s^6 -51s^5 + 
96s^4+44s^3+72s^2-12s - 32)}{m^2(s^2-2)^3(4s^3-3s^2-2)^2} \end{equation}

\begin{equation} S_{ab}S^{ab} = \frac{s^6(s+1)^2(s^2+2s -2)^2(3s^2-2s-2)^2
(5s^3+12s^2+6s+8)^2}{24m^2(s^2-2)^6(4s^3-3s^2-2)^4} \end{equation}

\noindent where $s \equiv \sin{\theta}$. The interesting thing is that 
at the point where this hypersurface crosses the horizon the latter quantity, and 
indeed the traceless Ricci scalar itself, vanishes: 

\begin{equation} s = \sqrt{3}-1 \hspace{5mm} \Rightarrow \hspace{5mm} 
S_{ab}S^{ab} = 0 \ . \label{S2} \end{equation}

\noindent There is a reason for this, as we will now explain. 

In many applications \cite{CFT} the near horizon limit of extremal 
black holes is of interest. Since this is not really our topic here we refer 
to the literature for the details in the Kerr case \cite{Bardeen}. The 
event horizon of the extremal black hole becomes a Killing horizon in 
the near horizon geometry, and the horizon Killing vector field is again 
null on a timelike hypersurface crossing this Killing horizon at a fixed 
latitude. Now it happens that a timelike hypersurface at fixed latitude 
$\theta$ in the near horizon geometry is in itself a 2+1 dimensional 
spacetime of considerable interest 
\cite{Sandin}. At high latitudes it is an anti-de Sitter space squashed 
along a spacelike fibre (along which it has also been made periodic). At 
low latitudes it is stretched along the same fibres. The Killing field that 
generates the horizon goes null also at latitude $\theta = \arcsin{(\sqrt{3}-1)}$. 
This hypersurface forms the boundary between squashing and stretching, and 
has the intrinsic geometry of a 2+1 dimensional anti-de Sitter space. 
This explains the result (\ref{S2}).  

\vspace{10mm}

{\bf 5. Conclusions}

\vspace{5mm}

\noindent The behaviour of the timelike Killing vector fields gives the 
extremal Kerr black hole an onion-like structure: in the equatorial plane 
the event horizon is surrounded by annuli in each of which some Killing vector field 
is timelike. Every point outside the horizon belongs to such an annulus, 
but none of the annuli reach all the way down to the horizon. 

The Kerr spacetime has multiple velocity-of-light surfaces where the horizon 
Killing field becomes null. One lies in the exterior, another lies inside 
the inner horizon and touches the singularity, and a third lies at 
negative values of $r$. As one approaches the extremal 
limit of the Kerr spacetime the former two merge, in such 
a way that the extremal velocity-of-light surface crosses the horizon at 
a fixed latitude $\theta \approx 47^{\circ}$. 

In the near horizon limit of the extremal black hole the velocity-of-light 
surface has the geometry of a 2+1 dimensional anti-de Sitter space (with 
one spacelike direction made periodic).

\

\

\noindent \underline{Acknowledgements}: We thank Istvan Racz, Jos\'e Senovilla, 
and Ted Jacobson for discussions. IB is 
supported by the Swedish Research Council under contract VR 621-2010-4060.

\end{document}